\documentclass[journal]{IEEEtran}
\usepackage[english]{babel}
\usepackage{graphicx}
\usepackage{amsmath}
\usepackage{mathtools}
\usepackage{cite}

\begin{document}

\title{Correlated Source Coded Sequences with Compromised Channel and Source Symbols using Shannon's Cipher System}
\author{R Balmahoon$^{*}$, A.J Han Vinck$^{\dag}$ and L Cheng$^{*}$
\\
\authorblockA{$^{*}$School of Electrical and Information Engineering\\University of the Witwatersrand\\Private Bag 3, Wits. 2050, Johannesburg, South Africa\\
Email: reevana.balmahoon@students.wits.ac.za, ling.cheng@wits.ac.za \\ $^{\dag}$ Institute for Experimental Mathematics\\Duisburg-Essen University, Ellernstr.29, 45328 Essen, Germany\\
Email: vinck@iem.uni-due.de}}

\maketitle

\begin{abstract}

Correlated sources are present in communication systems where protocols ensure that there is some predetermined information for sources to transmit. Here, two correlated sources across a channel with eavesdroppers are investigated, and conditions for perfect secrecy when some channel information and some source data symbols (the predetermined information) have been wiretapped are determined. The adversary in this situation has access to more information than if a link is wiretapped only and can thus determine more about a particular source. This scenario caters for an application where the eavesdropper has access to some preexisting information. We provide bounds for the channel and key rates for this scenario. Further, we provide a method to reduce the key lengths required for perfect secrecy.

\end{abstract}

%%%%%%%%%%%%%%%%%%%%%%%%%%%%%%%%%%%%%%%%%%%%%%%%%%%%%%%%%%%%%%%%%%%%%%%%%%%%%%%%
\section{INTRODUCTION}

Practical communication systems make use of correlated sources, for example smart grid meters. Each smart grid meter for a particular grid conforms to certain protocols and this means that certain information (e.g. date, area, etc.) in the header files will be the same for various meters. From the receiver's (or an eavesdropper's) perspective, it appears as common information shared between the meters. This is therefore pre-existing or known information for an eavesdropper. Thus, correlated source scenarios exist in abundance in systems transmitting information as dictated by communication protocols, e.g. smart grid meter systems. This implies that the theory used for correlated sources may also be applied to these types of systems. 

Correlated source coding incorporates the lossless compression of two or more correlated data streams. It has the ability to decrease the bandwidth required to transmit and receive messages because a compressed form of the original message is sent across the communication links instead of the original message. A compressed message has more information per bit, and therefore has a higher entropy because the transmitted information is more unpredictable. The unpredictability of the compressed message is also beneficial for the information security. Correlated source coding has come about to negate the security defect that is experienced when transmitting common information across channels; if one source's common information is known and it is known that the source is correlated to another source then information abotu the second source is also known. 

In practical communication systems links are prone to eavesdropping and as such this work incorporates eavesdropped links. In work by Aggarwal \textit{et al.} \cite{active_eavesdropper_aggarwal} it is seen that an eavesdropper can be active and can erase/modify bits. They develop a perfect secrecy model for this scenario. The eavesdropper that we investigate is a passive wiretapper, who cannot modify information. A wiretap channel has been used to indicate a scenario with an eavesdropper. The mathematical model for a wiretap channel has been given by Rouayheb \textit{et al.}~\cite{ref12_rouayheb_soljanin}, and can be explained as follows: the channel between a transmitter and receiver  is error-free and can transmit $n$ symbols from which $\mu$ of them can be observed by the eavesdropper and the maximum secure rate can be shown to equal $n-\mu$ symbols. The wiretap channel II was described by Ozarow and Wyner \cite{ref15_ozarow_wyner} with a coset coding scheme. A variation of the Gaussian wiretap channel has been investigated by Mitrpant \textit{et al.} \cite{gaussian_wiretap_mitrpant}. Characteristics on this channel were introduced by Luo \textit{et al.} \cite{ref14_luo_mitpant}. The characteristics focused on were those pertaining to Hamming weights and Hamming distances in order to determine the equivocation of a wiretapper. Thereafter, the security aspect of wiretap networks has been looked at in various ways by Cheng \textit{et al.} \cite{ref21_cheng_yeung}, and Cai and Yeung \cite{ref11_cai_yeung}, emphasizing that it is of concern to secure this type of channel. 

The difference between the original wiretap channel and the wiretap channel II is that the latter is error free. In an interesting application of the wiretap channel and wiretap channel of type II, Dai \textit{et al.} \cite{side_information_dai} presented a model that incorporates compromised encoded bits and wiretapped bits from a noisy channel. Here, we consider a scenario where an eavesdropper has access to more than just the bits from the communication links. Luo \textit{et al.} \cite{ref14_luo_mitpant}, in some previous work, have described this sort of adversary as more powerful. In addition to the eavesdropped bits from the communication links, the eavesdropper also has access to some data symbols from the source. In other previous work \cite{arxiv1_bal_ling}, the information leakage for two correlated sources when some channel information from the communication links had been wiretapped was investigated. The information leakage for the model described in Section II has also previously been analyzed \cite{allerton}. Intuitively from this work, it is seen that there is indeed more information gained by the more powerful eavesdropper. This makes it easier for the eavesdropper to determine the transmitted message and information about the alternate source.

This extra information that the eavesdropper has access to can be considered as side information to assist with decoding. Villard and Piantanida \cite{pablo_secure_multiterminal} have also looked at correlated sources: A source sends information to the receiver and an eavesdropper has access to information correlated to the source, which is used as side information. There is a second encoder that sends a compressed version of its own correlation observation of the source privately to the receiver. Here, the authors show that the use of correlation decreases the required communication rate and increases secrecy. Villard \textit{et al.} \cite{pablo_secure_transmission_receivers} have explored this side information concept further where security using side information at the receiver and eavesdropper is investigated. Side information is generally used to assist the decoder to determine the transmitted message. An earlier work involving side information was done by Yang \textit{et al.}~\cite{feedback_yang}. The concept can be considered to be generalized in that the side information could represent a source. It is an interesting problem when one source is more important and Hayashi and Yamamoto\cite{Hayashi_coding} have considered it in another scheme with two sources, where only one source is secure against wiretappers and the other must be transmitted to a legitimate receiver. They develop a security criterion based on the number of correct guesses of a wiretapper to attain a message. In this paper the source data symbols may be seen as side information to the eavesdropper, which is depicted in Section II.

In cryptographic systems, there is usually a message in plaintext that needs to be sent to a receiver. In order to secure it, the plaintext is encrypted so as to prevent eavesdroppers from reading its contents. This ciphertext is then transmitted to the receiver. Shannon's cipher system (mentioned by Yamamoto\cite{shannon1_yamamoto}) incorporates this idea. The definition of Shannon's cipher system has been discussed by Hanawal and Sundaresan~\cite{hanawal_shannon}. In Yamamoto's~\cite{shannon1_yamamoto} development on this model, a correlated source approach is introduced. 

Shannon's secrecy model is an interesting avenue for this work. Previous work \cite{arxiv1_bal_ling} has looked at a model for Shannon's cipher system when there is an eavesdropper at the channel only. Merhav \cite{shannon_secrecy_merhav} investigated similarly, for a model using the additional parameters; namely, the distortion of the source reconstruction at the legitimate receiver, the bandwidth expansion factor of the coded channels, and the average transmission cost.

The paper is arranged in four sections. Section II puts forth the model and its description, where the proofs for the channel and key rate bounds to attain perfect secrecy for this model described here are detailed in Section IV. Section III details a method to reduce the key length. Lastly, the paper is concluded in Section V.

\section{Model}

\begin{figure}[ht]
\centering
\includegraphics [scale = 0.7]{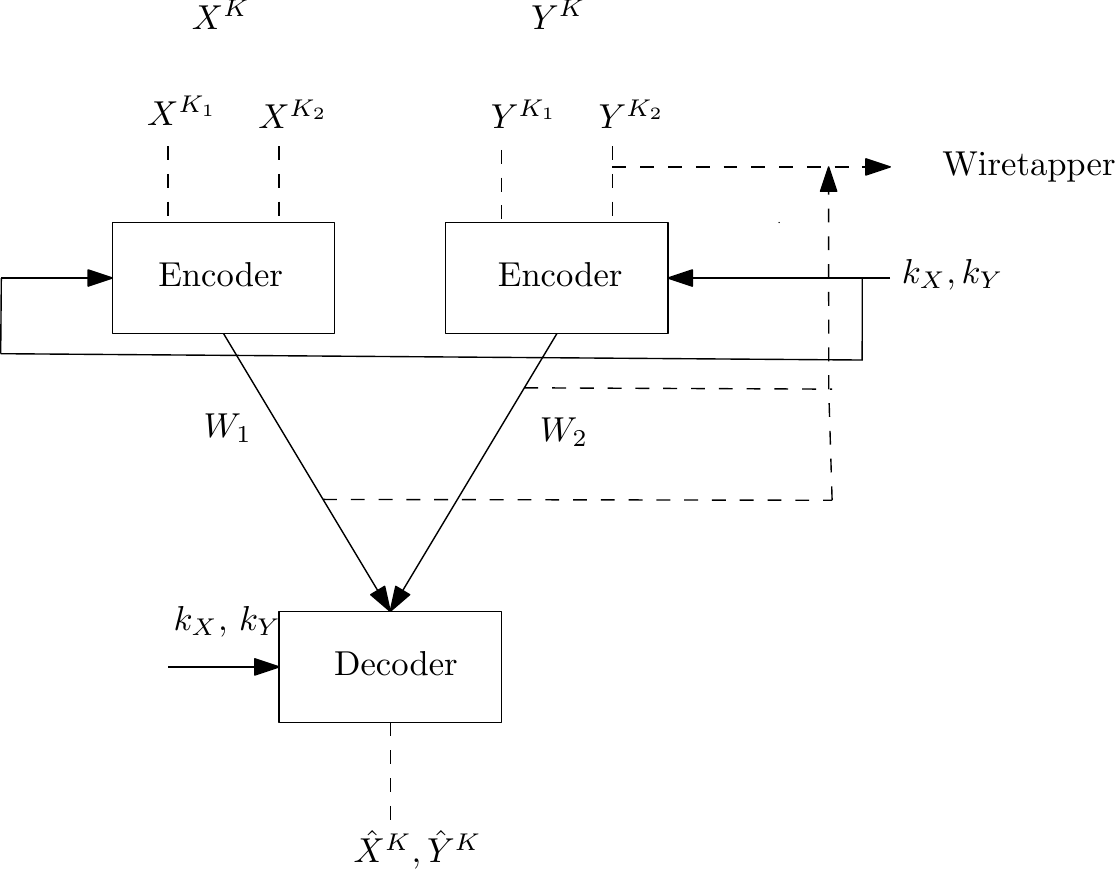}
\caption{Shannon cipher system for two correlated sources with wiretapped source symbols}
\label{fig:shannon_cipher_2sources}
\end{figure}

\begin{figure}[ht]
\centering
\includegraphics [scale = 0.7]{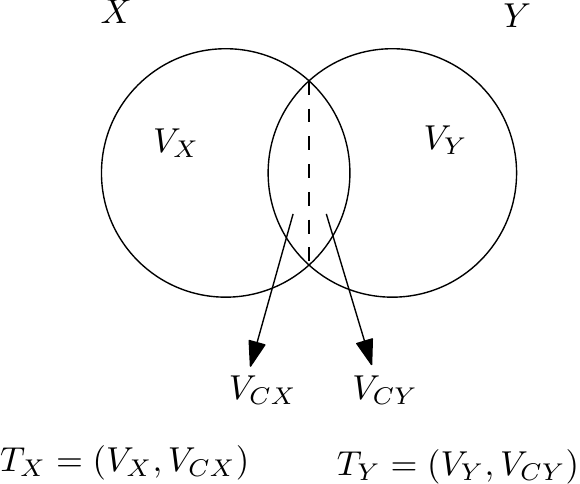}
\caption{The relation between private and common information}
\label{fig:new_venn2}
\end{figure}

The independent, identically distributed (i.i.d.) sources $X$ and $Y$ are mutually correlated random variables, depicted in Figure~\ref{fig:shannon_cipher_2sources}. The alphabet sets for sources $X$ and $Y$ are represented by $\mathcal{X}$ and $\mathcal{Y}$ respectively. Assume that ($X^K$, $Y^K$) are encoded into two channel information portions ($T_{X}$ and $T_{Y}$). We can write $T_X = F(X^K)$ and $T_Y = F(Y^K)$ where $T_X$ and $T_Y$ are the channel information of $X$ and $Y$. This channel information can be represented by common or private information. Here, $T_X$ and $T_Y$ are characterized by $(V_X, V_{CX}) = F'(T_X)$ and $(V_Y, V_{CY}) = F'(T_Y)$. The Venn diagram in Figure \ref{fig:new_venn2}  easily illustrates this idea where it is shown that $V_X$ and $V_Y$ represent the private information of sources $X$ and $Y$ respectively and $V_{CX}$ and $V_{CY}$ represent the common information between $X^K$ and $Y^K$ generated by $X^K$ and $Y^K$ respectively. Each source is composed of two components; $X^{K_1}$ and $X^{K_2}$ for $X^K$ and $Y^{K_1}$ and $Y^{K_2}$ for $Y^K$, of which one component is leaked to the eavesdropper. Here, the lengths $K_1$ and $K_2$ are related to $K$ as follows: $K_1 + K_2 = K$. Due to the stationary nature of the sources, if $Y^{K_2}$ is known by the wiretapper then it corresponds to $X^{K_2}$ been known about $X^K$, as the wiretapper has access to certain common information between the sources. 

Both the transmitter and receiver have access to the key, a random variable, independent of $X^K$ and $Y^K$ and taking values in $I_{M_K} = \{0, 1, 2, \ldots ,M_{K} - 1\}$. The sources $X^K$ and $Y^K$ compute the ciphertexts $X^{'}$ and $Y^{'}$, which are the result of specific encryption functions on the plaintext from $X$ and $Y$ respectively. The encryption functions are invertible, thus knowing $X^{'}$ and the key, $X^K$ can be retrieved. The mutual information between the plaintext and ciphertext should be small so that the wiretapper cannot gain much information about the plaintext. For perfect secrecy, this mutual information should be zero, then the length of the key should be at least the length of the plaintext.

The correlated sources $X$ and $Y$ transmit messages (in the form of some channel information) to the receiver along the channel with an eavesdropper. The decoder determines $X$ and $Y$ only after receiving all of $T_X$ and $T_Y$. The common information between the sources are transmitted through the portions $V_{CX}$ and $V_{CY}$. In order to decode a transmitted message, a source's private information and both common information portions are necessary. This aids in security as it is not possible to determine, for example $X$ by wiretapping all the contents transmitted along $X$'s channel and the part of $X$'s data symbols corresponding to the wiretapped channel information only. This is different to Yamamoto's~\cite{shannon1_yamamoto} model as here the common information consists of two portions. 

The eavesdropper has access to either the common or private portions given by $T_Y$ and/or $T_X$ and some data symbols from the corresponding source ($Y^{K_2}$). The effect is that the eavesdropper has access to some compressed information (that is transmitted across the communication link after encoding) and some uncompressed information (i.e. the source's data symbols). There is a mapping/function that describes the relation between the uncompressed information and the compressed information. This implies that certain source bits correspond to certain compressed bits transmitted as channel information. It is valuable to determine how much of information to transmit at a time such that the eavesdropper cannot access the information (this is described in the next section).

The correlated sources $X$ and $Y$ transmit messages (in the form of syndromes) to the receiver along wiretapped links. The decoder determines $X$ and $Y$ only after receiving all of $T_X$ and $T_Y$. The common information between the sources are transmitted through the portions $V_{CX}$ and $V_{CY}$. In order to decode a transmitted message, a source's private information and both common information portions are necessary. This aids in security as it is not possible to determine, for example $X$ by wiretapping all the contents transmitted along $X$'s channel only. This is different to Yamamoto's~\cite{shannon1_yamamoto} model as here the common information consists of two portions.  The aim is to keep the system as secure as possible and these following sections show how it is achieved by this new model.

The encoder functions for $X$ and $Y$, ($E_X$ and $E_Y$ respectively) are given as:

\begin{eqnarray}
E_X : (\mathcal{X}^{K_1}, \mathcal{X}^{K_2}) \times I_{M_{kX}} & \rightarrow & I_{M_X'} =  \{0, 1, \ldots, M_X' - 1\} \nonumber 
\\ && I_{M_{CX}'} =  \{0, 1, \ldots, 
\\ && M_{CX}' - 1\}
\label{xencoder_fcn}
\end{eqnarray}

\begin{eqnarray}
E_Y : (\mathcal{Y}^{K_1}, \mathcal{Y}^{K_2}) \times I_{M_{kY}} & \rightarrow & I_{M_Y'} =  \{0, 1, \ldots, M_Y' - 1\} \nonumber 
\\ && I_{M_{CY}'} =  \{0, 1, \ldots, 
\\ && M_{CY}' - 1\}
\label{yencoder_fcn}
\end{eqnarray}

The decoder is defined as:

\begin{eqnarray}
D_{XY} : (I_{M'_X}, I_{M'_Y}, I_{M'_{CX}},I_{M'_{CY}})  & \times &  I_{M_{kX}}, I_{M_{kY}} \nonumber \\
& \rightarrow & \mathcal{X}^K \times \mathcal{Y}^K
\end{eqnarray}

The encoder and decoder mappings are below:
\begin{eqnarray}
W_1 = F_{E_X} (X^{K_1}, X^{K_2}, W_{kX})
\end{eqnarray}

\begin{eqnarray}
W_2 = F_{E_Y} (Y^{K_1}, Y^{K_2}, W_{kY})
\end{eqnarray}

\begin{eqnarray}
\widehat{X}^K = F_{D_X} (W_1, W_2, W_{kX})
\end{eqnarray}

\begin{eqnarray}
\widehat{Y}^K = F_{D_Y} (W_1, W_2, W_{kY})
\end{eqnarray}

or 

\begin{eqnarray}
(\widehat{X}^K, \widehat{Y}^K) = F_{D_{XY}} (W_1, W_2, W_{kX}, W_{kY})
\end{eqnarray}

The following conditions should be satisfied for cases 1- 4:

\begin{eqnarray}
\frac{1}{K}\log M_X \le R_X +\epsilon
\label{cond1}
\end{eqnarray}

\begin{eqnarray}
\frac{1}{K}\log M_Y \le R_Y +\epsilon
\label{cond2}
\end{eqnarray}

\begin{eqnarray}
\frac{1}{K}\log M_{kX} \le R_{X} +\epsilon
\label{cond3}
\end{eqnarray}

\begin{eqnarray}
\frac{1}{K}\log M_{kY} \le R_{{kY}} +\epsilon
\label{cond4}
\end{eqnarray}

\begin{eqnarray}
\text {Pr} \{\widehat{X}^K \neq X^K\} \le \epsilon
\label{cond5}
\end{eqnarray}

\begin{eqnarray}
\text{Pr} \{ \widehat{Y}^K \neq Y^K\} \le \epsilon
\label{cond6}
\end{eqnarray}

\begin{eqnarray}
\frac{1}{K} H(X^K|W_1, W_2) \le h_X + \epsilon
\label{cond7}
\end{eqnarray}

\begin{eqnarray}
\frac{1}{K} H(Y^K|W_1, W_2) \le h_Y + \epsilon
\label{cond8}
\end{eqnarray}

\begin{eqnarray}
\frac{1}{K} H(X^K,Y^K|W_1, W_2) \le h_{XY} + \epsilon
\label{cond8.1}
\end{eqnarray}

where $R_X$ is the rate of source $X$'s channel and $R_Y$ is the rate of source $Y$'s channel. Here, $(R_{kX}, R_{kY})$ is the rate of the key channel when allocating a key to $X$ and $Y$. The security level for $X$ and $Y$ are measured by the total and individual uncertainties, $(h_X, h_Y)$. Here, $h_{XY}^X$ and $h_{XY}^Y$ represent the security level of $X$ and $Y$ with respect to $X$ and $Y$ respectively.
\\\\
The cases 1 - 3 are:
\\ \textit{Case 1:} When $(W_1, W_2, Y^{K_2})$ is leaked and $(X^K, Y^K)$ needs to be kept secret. The security level of concern is represented by $\frac{1}{K}H(X^K, Y^K|W_1, W_2)$.
\\ \textit{Case 2:} When $(W_1, W_2, Y^{K_2})$ is leaked and $(X^K, Y^K)$ needs to be kept secret. The security level of concern is represented by $(\frac{1}{K}H(X^K|W_1, W_2), \frac{1}{K}H(Y^K|W_1, W_2))$.
\\ \textit{Case 3:} When $(W_1, W_2, Y^{K_2})$ is leaked and $Y^K$ needs to be kept secret.The security level of concern is represented by $\frac{1}{K}H(Y^K|W_1, W_2)$.
\\ 
\\\\
The admissible rate region for each case is defined as follows:
\\ \textit{Definition 1a:} ($R_X$, $R_Y$, $R_{kX}$, $R_{kY}$, $h_{XY}$) is admissible for case 1 if there exists a code ($F_{E_{X}}$, $F_{D_{XY}}$) such that \eqref{cond1} - \eqref{cond6} and \eqref{cond8.1} hold for any $\epsilon \rightarrow 0$ and sufficiently large $K$.
\\ \textit{Definition 1b:} ($R_X$, $R_Y$, $R_{kX}$, $R_{kY}$, $h_X$,  $h_{Y}$) is admissible for case 2 if there exists a code ($F_{E_{Y}}$, $F_{D_{XY}}$) such that \eqref{cond1} - \eqref{cond8} hold for any $\epsilon \rightarrow 0$ and sufficiently large $K$.
\\ \textit{Definition 1c:} ($R_X$, $R_Y$, $R_{kX}$, $R_{kY}$, $h_{Y}$) is admissible for case 3 if there exists a code ($F_{E_{X}}$, $F_{D_{XY}}$) and ($F_{E_{Y}}$, $F_{D_{XY}}$) such that \eqref{cond1} - \eqref{cond6} and \eqref{cond8} hold for any $\epsilon \rightarrow 0$ and sufficiently large $K$.
\\ \textit{Definition 2:} The admissible rate regions of $\mathcal{R}_j$ for case $j$ are defined as:

\begin{eqnarray}
\mathcal{R}_1(h_{XY}) = \{(R_X, R_Y, R_{kX}, R_{kY}):			\nonumber
\\(R_X, R_Y, R_{kX}, R_{kY}, h_{XY} ) \text{ is admissible for case 1} \}
\end{eqnarray}

\begin{eqnarray}
\mathcal{R}_2(h_X, h_{Y}) = \{(R_X, R_Y, R_{kX}, R_{kY}):			\nonumber
\\ (R_X, R_Y, R_{kX}, R_{kY},  h_{X}, h_{Y} ) \text{ is admissible for case 2} \}
\end{eqnarray}

\begin{eqnarray}
\mathcal{R}_3(h_{Y}) = \{(R_X, R_Y, R_{kX}, R_{kY}):			\nonumber
\\ (R_X, R_Y, R_{kX}, R_{kY}, h_{Y} ) \text{ is admissible for case 3} \}
\end{eqnarray}

Theorems for these regions have been developed:

\textit{Theorem 1:} For $0 \le h_{XY} \le H(X,Y) - \mu_C -\mu_Y$ and
\begin{eqnarray}
&& \mathcal{R}_{1}(h_{XY}) = \{(R_X, R_Y, R_{kX},R_{kY}): 		\nonumber
\\ && R_X \geq H(X|Y), 				\nonumber
\\ && R_Y \geq H(Y|X),			\nonumber
\\ && R_X + R_Y \geq H(X,Y) 		\text { and } 						\nonumber
\\ && R_{kX} + R_{kY} \geq h_{XY}   \}			
\label{theorem2}
\end{eqnarray}

\textit{Theorem 2:} For $0 \le h_{X} \le H(X) - \mu_C$  and 
\\ $0  \le h_{Y} \le H(Y) - \mu_C - \mu_Y$,
\begin{eqnarray}
&& \mathcal{R}_{2}(h_X, h_{Y}) = \{(R_X, R_Y, R_{kX},R_{kY}): 		\nonumber
\\ && R_X \geq H(X|Y), 				\nonumber
\\ && R_Y \geq H(Y|X),			\nonumber
\\ && R_X + R_Y \geq H(X,Y) 	\text { and } 						\nonumber
\\ && R_{kX} + R_{kY} \geq \text{max}(h_{X}, h_Y)   \}	
\label{theorem3}
\end{eqnarray}

where $\mathcal{R}_{1}$ and $\mathcal{R}_{2}$ are the regions for cases 1 and 2 respectively. 
\\
Corollary 1:
For $0 \le h_Y \le H(Y) - \mu_C - \mu_Y$,
\\$\mathcal{R}_{3} (h_Y) = \mathcal{R}_{2}(0, h_Y)$
\\
The code achieving these bounds for case 2 allows for the key length of $h_Y$ to be achieved across $Y$'s channel in case 3. 
The security levels, which are measured by the individual uncertainties $(h_X, h_Y)$ and total uncertainty $h_{XY}$ give an indication of the level of uncertainty in knowing certain information. When the uncertainty increases then less information is known to an eavesdropper and there is a higher level of security.

\section{Discussion}
Each message transmitted requires a key, as described in Section II. If only $Y$'s channel is wiretapped then $X$ does not need a key for keeping messages confidential. For this model however, we describe cases where the channel information from both $X$ and $Y$ are wiretapped. In all cases, $Y^{K_2}$ is also leaked to the wiretapper. This is what makes the adversary more powerful. This model may be used in a communication system that has some preexisting information known. This is common as information known by certain communication protocols could serve as the preexisting information. Here, $Y_{K_2}$ could contain some private information and some common information. The wiretapper could thus gain private and/or common information about the source $Y^K$. The information gained through $Y^{K_2}$ may or may not correspond to the information leaked across $Y$'s channel (the compressed and uncompressed information have a certain relationship, as described in Section II).

In this model, a key can be provided by either $X$ or $Y$ as we have a common key channel. This means that if $X$ allocates a key then $Y$ can make use of the same key once. 

Correlated source compression reduces the bandwidth of the transmitted message, hence the key length required for this scenario is less than a case for when there is no source compression. The reduced bandwidth is also a security advantage as the message entropy is higher. One masking method to reduce the key length is to use a portion where a key has already been allocated to mask a portion that needs to be covered. This is able to provide confidentiality as a key would.

\section{Proof of Theorems 1  - 2}

We construct a code based on the prototype code ($W_X, W_Y, W_{CX}, W_{CY}$) described below:
\\
For any $\epsilon_0 \geq 0$ and sufficiently large $K$, there exits a code $W_X = F_X(X^K)$, $W_Y = F_Y(Y^K)$, $W_{CX} = F_{CX}(X^K)$, $W_{CY} = F_{CY}(Y^K)$, $\widehat{X}^K,\widehat{Y}^K = G(W_X, W_Y, W_{CX}, W_{CY})$, where $W_X \in I_{M_X}$, $W_Y \in I_{M_Y}$, $W_{CX} \in I_{M_{CX}}$, $W_{CY} \in I_{M_{CY}}$ for $I_{M_{\alpha}}$, which is defined as $\{0, 1, \ldots, M_{\alpha} - 1\}$, that satisfies,

\begin{eqnarray}
Pr \{\widehat{X}^K, \widehat{Y}^K \neq X^K, Y^K\} \le \epsilon
\label{lemma1_1}
\end{eqnarray}

\begin{eqnarray}
H(X|Y) - \epsilon_0 & \le & \frac{1}{K} H(W_X) \le \frac{1}{K} \log M_X  \nonumber\\
& \le & H(X|Y) + \epsilon_0
\label{lemma1_2}
\end{eqnarray}

\begin{eqnarray}
H(Y|X) - \epsilon_0 & \le & \frac{1}{K} H(W_Y) \le \frac{1}{K} \log M_Y  \nonumber\\
& \le & H(Y|X) + \epsilon_0
\label{lemma1_3}
\end{eqnarray}

\begin{eqnarray}
& & I(X;Y) - \epsilon_0 \le \frac{1}{K} (H(W_{CX}) + H(W_{CY})) \nonumber \\
    & \le & \frac{1}{K} (\log M_{CX} + \log M_{CY}) \le I(X;Y) + \epsilon_0
\label{lemma1_4}
\end{eqnarray}

\begin{eqnarray}
\frac{1}{K} H(X^K|W_Y) \geq H(X) - \epsilon_0
\label{lemma1_5}
\end{eqnarray}

\begin{eqnarray}
\frac{1}{K} H(Y^K|W_X) \geq H(Y) - \epsilon_0
\label{lemma1_6}
\end{eqnarray}

We can see that \eqref{lemma1_2} - \eqref{lemma1_4} mean
\begin{eqnarray}
&& H(X,Y) - 3\epsilon_0 \le \frac{1}{K} (H(W_X) + H(W_Y) + H(W_{CX}) \nonumber \\
& + & H(W_{CY})) \le H(X,Y) + 3\epsilon_0
\label{lemma1_7}
\end{eqnarray}

Hence from \eqref{lemma1_1}, \eqref{lemma1_7} and the ordinary source coding theorem, ($W_X$, $W_Y$, $W_{CX}$, $W_{CY}$) have no redundancy for sufficiently small $\epsilon_0 \geq 0$. It can also be seen that $W_X$ and $W_Y$ are independent of $Y^K$ and $X^K$ respectively. Equations \eqref{lemma1_1} - \eqref{lemma1_7} have been proven in \cite{arxiv1_bal_ling}.
In order to include a key in the prototype code, $W_X$ is divided into two parts as per the method used by Yamamoto \cite{shannon1_yamamoto}:
\begin{eqnarray}
W_{X1} = W_X \text{ mod } M_{X1} \in I_{M_{X1}} = \{0, 1, 2, \ldots, M_{X1} - 1\}
\label{theorems2-4_eq_1}
\end{eqnarray}

\begin{eqnarray}
W_{X2} = \frac{W_X - W_{X1}}{M_{X1}} \in I_{M_{X2}} = \{0, 1, 2, \ldots, M_{X2} - 1\}
\label{theorems2-4_eq_2}
\end{eqnarray}

where $M_{X1}$ is a given integer and $M_{X2}$ is the ceiling of $M_X/M_{X1}$. The $M_X/M_{X1}$ is considered an integer for simplicity, because the difference between the ceiling value and the actual value can be ignored when $K$ is sufficiently large. In the same way, $W_Y$ is divided:

\begin{eqnarray}
W_{Y1} = W_Y \text{ mod } M_{Y1} \in I_{M_{Y1}} = \{0, 1, 2, \ldots, M_{Y1} - 1\}
\label{theorems2-4_eq_3}
\end{eqnarray}

\begin{eqnarray}
W_{Y2} = \frac{W_Y - W_{Y1}}{M_{Y1}} \in I_{M_{Y2}} = \{0, 1, 2, \ldots, M_{Y2} - 1\}
\label{theorems2-4_eq_4}
\end{eqnarray}

The common information components $W_{CX}$ and $W_{CY}$ are already portions and are not divided further. In this scenario $W_{CX}+W_{CY}$ lies between $0$ and $I(X;Y)$. It can be represented by $X$ and $Y$, $X$ only or $Y$ only.
It can be shown that when some of the codewords are wiretapped the uncertainties of $X^K$ and $Y^K$ are bounded as follows:

\begin{eqnarray}
\frac{1}{K} H(X^K|W_{X2},W_Y) \geq I(X;Y) + \frac{1}{K} \log M_{X1} - \epsilon_{0}^{'}
\label{theorems2-4_ineq_1}
\end{eqnarray}

\begin{eqnarray}
\frac{1}{K} H(Y^K|W_{X},W_{Y2}) \geq I(X;Y) + \frac{1}{K} \log M_{Y1} - \epsilon_{0}^{'}
\label{theorems2-4_ineq_2}
\end{eqnarray}

\begin{eqnarray}
\frac{1}{K} H(X^K|W_{X},W_{Y2}) \geq I(X;Y) - \epsilon_{0}^{'}
\label{theorems2-4_ineq_3}
\end{eqnarray}

\begin{eqnarray}
\frac{1}{K} H(X^K|W_{X},W_Y, W_{CY}) \geq \frac{1}{K} \log M_{CX} - \epsilon_{0}^{'}
\label{theorems2-4_ineq_4}
\end{eqnarray}

\begin{eqnarray}
\frac{1}{K} H(Y^K|W_{X},W_Y, W_{CY}) \geq \frac{1}{K} \log M_{CX} - \epsilon_{0}^{'}
\label{theorems2-4_ineq_5}
\end{eqnarray}

\begin{eqnarray}
\frac{1}{K} H(X^K|W_Y, W_{CY}) \geq H(X|Y) + \frac{1}{K} \log M_{CX} - \epsilon_{0}^{'}
\label{theorems2-4_ineq_6}
\end{eqnarray}

\begin{eqnarray}
\frac{1}{K} H(Y^K|W_Y, W_{CY}) \geq H(Y|X) + \frac{1}{K} \log M_{CX} - \epsilon_{0}^{'}
\label{theorems2-4_ineq_7}
\end{eqnarray}

\begin{eqnarray}
\frac{1}{K_2} H(Y^{K_2}) \geq \mu_C + \mu_Y - \epsilon_{0}^{'}
\label{theorems2-4_ineq_7.1}
\end{eqnarray}

where $\epsilon_{0}^{'} \rightarrow 0$ as  $\epsilon_{0} \rightarrow 0$.
Here, we indicate the wiretapped source symbols with the entropy in \eqref{theorems2-4_ineq_7.1}, where $\mu_C$ and $\mu_Y$ are the common and private portions of the i.i.d source $Y$ that are contained in $Y^{K_2}$ per symbol. Here, $Y_{K_2}$ is proportional to $K$. 
\\
The proofs for \eqref{theorems2-4_ineq_1} - \eqref{theorems2-4_ineq_7} are the same as per Yamamoto's\cite{shannon1_yamamoto} proof in Lemma A1. The difference is that $W_{CX}$, $W_{CY}$, $M_{CX}$ and $M_{CY}$ are described as $W_{C1}$, $W_{C2}$, $M_{C1}$ and $M_{C2}$ respectively by Yamamoto. Here, we consider that $W_{CX}$ and $W_{CY}$ are represented by Yamamoto's $W_{C1}$ and $W_{C2}$ respectively. In addition there are some more inequalities considered here to describe the model better:
\begin{eqnarray}
 \frac{1}{K} H(Y^K|W_X, W_{CX}, W_{CY}, W_{Y2}) & \geq & \frac{1}{K} \log M_{Y1}  \nonumber
 \\ & - & \epsilon_{0}^{'}
\label{theorems2-4_ineq_8}
\end{eqnarray}

\begin{eqnarray}
 \frac{1}{K} H(Y^K|W_X, W_{CX}, W_{CY}) & \geq & \frac{1}{K} \log M_{Y1}  \nonumber
\\ & + & \frac{1}{K} \log M_{Y2} - \epsilon_{0}^{'}
\label{theorems2-4_ineq_9}
\end{eqnarray}

\begin{eqnarray}
\frac{1}{K} H(X^K|W_{X2}, W_{CY}) & \geq & \frac{1}{K} \log M_{X1} 	\nonumber
\\ & + & \frac{1}{K} \log M_{CX} - \epsilon_{0}^{'}
\label{theorems2-4_ineq_10}
\end{eqnarray}

\begin{eqnarray}
\frac{1}{K} H(Y^K|W_{X2}, W_{CY}) & \geq & \frac{1}{K} \log M_{Y1} 	\nonumber
\\ & + & \frac{1}{K} \log M_{Y2} + \frac{1}{K} \log M_{CX} 			\nonumber
\\ & - & \epsilon_{0}^{'}
\label{theorems2-4_ineq_11}
\end{eqnarray}

The inequalities \eqref{theorems2-4_ineq_8} and \eqref{theorems2-4_ineq_9} can be proved in the same way as per Yamamoto's\cite{shannon1_yamamoto} Lemma A2, and  \eqref{theorems2-4_ineq_10} and \eqref{theorems2-4_ineq_11} can be proved in the same way as per Yamamoto's\cite{shannon1_yamamoto} Lemma A1.

%%%%%%%%%%%%%%%%%%%%%%%%%%%%%%%%%%%

\begin{proof}[Proof of Theorem 1]
Suppose that ($R_X$, $R_Y$, $R_{kX}$, $R_{kY}$) $\in$ 
$\mathcal{R}_{1}$ for $h_{XY} \le H(X,Y) - \mu_C - \mu_Y$. Without loss of generality, we assume that $h_{XY} \le R_{kX} + R_{kY}$. Then, from 
\eqref{theorem2} 
\begin{eqnarray}
&& R_X \geq H(X^K|Y^K)  				\nonumber
\\&& R_Y  \geq H(Y^K|X^K) 				\nonumber
\\&& R_X + R_Y  \geq H(X^K, Y^K)
\label{theorem1_proof_1}
\end{eqnarray}

\begin{eqnarray}
R_{kX} + R_{kY} \geq h_{XY}
\label{theorem1_proof_2}
\end{eqnarray}

Here the keys are uniform random numbers. For the first case, consider the following: $h_{XY} > I(X;Y)$.

\begin{eqnarray}
M_{X1} = \text{min}(2^{K H(X|Y)}, 2^{K (h_{XY} - I(X;Y))})
\label{theorem1_proof_4.4}
\end{eqnarray}

\begin{eqnarray}
M_{Y1} = 2^{K (h_{XY} - I(X;Y))}
\label{theorem1_proof_4.6}
\end{eqnarray}

The codewords $W_1$ and $W_2$ and the key $W_{kX}$and $W_{kY}$ are now defined:

\begin{eqnarray}
W_1 = (W_{X1} \oplus W_{kY1},  W_{X2}, W_{CX} \oplus W_{kCX}) 
\label{theorem1_proof_7}
\end{eqnarray}

\begin{eqnarray}
W_2 = (W_{Y1} \oplus W_{kY1}, W_{Y2}, W_{CY} \oplus W_{kCY})
\label{theorem1_proof_8}
\end{eqnarray}

\begin{eqnarray}
W_{kX} = W_{kCX}
\label{theorem1_proof_9}
\end{eqnarray}

\begin{eqnarray}
W_{kY} = (W_{kY1}, W_{kCY})
\label{theorem1_proof_10}
\end{eqnarray}

where $W_\alpha \in I_{M_\alpha} = \{0, 1, \ldots, M_\alpha - 1\}$. The wiretapper will not know $W_{X1}$, $W_{CX}$ $W_{Y1}$ and $W_{CY}$ as these are protected by keys.

In this case, $R_X$, $R_Y$, $R_{kX}$ and $R_{kY}$ satisfy from \eqref{lemma1_2} - \eqref{lemma1_4} and \eqref{theorem1_proof_1} - \eqref{theorem1_proof_10}, that

\begin{eqnarray}
\frac{1}{K} \log M_X + \frac{1}{K} \log M_Y  & = & \frac{1}{K} (\log M_{X1} + \log M_{X2}  \nonumber
\\ & + &  \log M_{CX}) + \frac{1}{K} (\log M_{Y1}  \nonumber
\\ & + &  \log M_{Y2} +  \log M_{CY}) \nonumber 
\\ & \le & H(X|Y) + H(Y|X) 				\nonumber
\\ & + & I(X;Y) + 3 \epsilon_0 \nonumber
\\ & = & H(X,Y) + 3 \epsilon_0 		\nonumber
\\ & \le & R_X + R_Y + 3 \epsilon_0
\label{theorem1_proof_11}
\end{eqnarray}

\begin{eqnarray}
&& \frac{1}{K} [\log M_{kX} + \log M_{kY}]  \nonumber
\\ & = & \frac{1}{K} [\log M_{CX} + \log M_{CY} + \log M_{Y1}] 	\nonumber	
\\ & \le & I(X;Y) + h_{XY} - I(X;Y) -\epsilon_0  \label{num3.2.1}
\\ & = & h_{XY} - \epsilon_0		\nonumber
\\ & \le & R_{kX} + R_{kY} - \epsilon_0
\label{theorem1_proof_13}
\end{eqnarray}

where \eqref{num3.2.1} results from \eqref{theorem1_proof_4.6}.

The security levels thus result:
\begin{eqnarray}
&& \frac{1}{K} H(X^K, Y^K|W_1, W_2, Y^{K_2}) \nonumber
\\ & = & \frac{1}{K} H(X^K, Y^K|W_{X1} \oplus W_{kY1}, \nonumber
\\ && W_{X2}, W_{CX} \oplus W_{kCX}		\nonumber
\\ && W_{Y1} \oplus W_{kY1}, W_{Y2} \nonumber
\\ && W_{CY} \oplus W_{kCY}, Y^{K_2})			\nonumber
\\ & \ge & \frac{1}{K} H(X^K, Y^K|W_{X1}, W_{X2}, \nonumber
\\ && W_{Y1} \oplus W_{kY1}, W_{Y2}, Y^{K_2}) - \epsilon_0^{''}			\label{num5}
\\ & = & \frac{1}{K} H(X^K, Y^K|W_{X}, W_{Y2}, Y^{K_2}) - \epsilon_0^{''}		\nonumber
\\ & \geq & I(X;Y) + \frac{1}{K} \log M_{Y1} \nonumber
\\ & - & \mu_C - \mu_Y - 2\epsilon_0^{'} - \epsilon_0^{''}		\nonumber
\\ & = & I(X;Y)+ h_{XY} - I(X;Y)  - \mu_C - \mu_Y  \nonumber
\\ & - & 2\epsilon_0^{'} - \epsilon_0^{''}		\nonumber
\\ & = & h_{XY} - \mu_C - \mu_Y- 2\epsilon_0^{'} - \epsilon_0^{''}
\label{theorem1_proof_16}
\end{eqnarray}

where \eqref{num5} holds because $W_{CX}$ and $W_{CY}$ are covered by uniform random keys and the result of Yamamoto's Lemma A2.

Therefore ($R_X$, $R_Y$, $R_{kX}$, $R_{kY}$, $h_{XY}$) is admissible from \eqref{theorem1_proof_11} -  \eqref{theorem1_proof_16}.

Next the case where: $h_{XY} \le I(X;Y)$ is considered. 
The codewords and keys are now defined:

\begin{eqnarray}
W_1 = (W_{X1},  W_{X2}, W_{CX} \oplus W_{kCX}) 
\label{theorem1_proof_7.1.1}
\end{eqnarray}

\begin{eqnarray}
W_2 = (W_{Y1}, W_{Y2}, W_{CY})
\label{theorem1_proof_8.1.1}
\end{eqnarray}

\begin{eqnarray}
W_{kX} = (W_{kCX})
\label{theorem1_proof_10.1.1}
\end{eqnarray}

\begin{eqnarray}
M_{CX} = 2^{K h_{XY}}
\label{theorem1_proof_9.1.33}
\end{eqnarray}

where $W_\alpha \in I_{M_\alpha} = \{0, 1, \ldots, M_\alpha - 1\}$. The wiretapper will not know the $W_X$ and $W_Y$ that are covered with keys.

In this case, $R_X$, $R_Y$, $R_{kX}$ and $R_{kY}$ satisfy that

\begin{eqnarray}
\frac{1}{K} [\log M_{kX} + \log M_{kY}] & = & \frac{1}{K} \log M_{CX}	\nonumber
\\ & = & h_{XY} \nonumber \label{num333.1}
\\ & \le & R_{kX} + R_{kY} 
\label{theorem1_proof_13.1.1}
\end{eqnarray}

where \eqref{num333.1} results from \eqref{theorem1_proof_9.1.33}.

The security level thus results:
\begin{eqnarray}
\frac{1}{K} H(X^K, Y^K|W_1, W_2, Y^{K_2}) & = & \frac{1}{K} H(X^K, Y^K|W_{X1}, W_{X2}, \nonumber
\\ && W_{CX} \oplus W_{kCX}, \nonumber
\\ && W_{Y1}, W_{Y2},	W_{CY},	\nonumber
\\ && Y^{K_2})			\nonumber
\\ & \geq & \frac{1}{K} \log M_{CX} - \mu_C - 2\epsilon_0^{'}		\nonumber
\\ & = & h_{XY} - \mu_C - 2\epsilon_0^{'}  \label{num5.1.1}
\\ & \geq & h_{XY} - 2\epsilon_0^{'}
\label{theorem1_proof_16.1.1}
\end{eqnarray}

where \eqref{num5.1.1} holds from \eqref{theorem1_proof_9.1.33}.

Therefore ($R_X$, $R_Y$, $R_{kX}$, $R_{kY}$, $h_{XY}$) is admissible from \eqref{theorem1_proof_7.1.1} -  \eqref{theorem1_proof_16.1.1}.

\end{proof}

\begin{proof}[Proof of Theorem 2]
In the same way, Suppose that ($R_X$, $R_Y$, $R_{kX}$, $R_{kY}$) $\in$ 
$\mathcal{R}_{2}$ for $h_{X} \le H(X) - \mu_C$ and $h_Y \le H(Y) - \mu_C - \mu_Y$. Without loss of generality, we assume that $h_X \le h_Y$ and  $h_X + h_Y \le R_{kX} + R_{kY}$. Then, from \eqref{theorem3} 
\begin{eqnarray}
&& R_X \geq H(X^K|Y^K)  				\nonumber
\\&& R_Y  \geq H(Y^K|X^K) 				\nonumber
\\&& R_X + R_Y  \geq H(X^K, Y^K)
\label{theorem2_proof_1}
\end{eqnarray}

\begin{eqnarray}
h_X + h_Y \le R_{kX} + R_{kY}
\label{theorem2_proof_2}
\end{eqnarray}

Consider the following: $h_X > I(X;Y)$.

\begin{eqnarray}
M_{X1} = \text{min}(2^{K H(X|Y)}, 2^{K(h_Y - I(X;Y))})
\label{theorem2_proof_61}
\end{eqnarray}

\begin{eqnarray}
M_{Y1} = 2^{K (h_Y - I(X;Y))}
\label{theorem1_proof_65}
\end{eqnarray}

The codeword $W_2$ and the key $W_{kY}$ is now defined:

\begin{eqnarray}
W_1 = (W_{X1} \oplus W_{kY1},  W_{X2}, W_{CX} \oplus W_{kCX}) 
\label{theorem2_proof_81}
\end{eqnarray}

\begin{eqnarray}
W_2 = (W_{Y1} \oplus W_{kY1}, W_{Y2}, W_{CY} \oplus W_{kCY})
\label{theorem1_proof_82}
\end{eqnarray}

\begin{eqnarray}
W_{kX} = W_{kCX}
\label{theorem2_proof_83}
\end{eqnarray}

\begin{eqnarray}
W_{kY} = (W_{kY1}, W_{kCY})
\label{theorem1_proof_84}
\end{eqnarray}

In this case, $R_X$, $R_Y$, $R_{kX}$ and $R_{kY}$ satisfy from \eqref{lemma1_2} - \eqref{lemma1_4} and \eqref{theorem2_proof_61} - \eqref{theorem1_proof_84}, that

\begin{eqnarray}
\frac{1}{K} \log M_X + \frac{1}{K} \log M_Y  & = & \frac{1}{K} (\log M_{X1} + \log M_{X2}  \nonumber
\\ & + &  \log M_{CX}) + \frac{1}{K} (\log M_{Y1}  \nonumber
\\ & + &  \log M_{Y2} +  \log M_{CY}) \nonumber 
\\ & \le & H(X|Y) + H(Y|X) 				\nonumber
\\ & + & I(X;Y) + 3 \epsilon_0 \nonumber
\\ & = & H(X,Y) + 3 \epsilon_0 		\nonumber
\\ & \le & R_X + R_Y + 3 \epsilon_0
\label{theorem2_proof_11}
\end{eqnarray}

\begin{eqnarray}
&& \frac{1}{K} [\log M_{kX} + \log M_{kY}]  \nonumber
\\ & = & \frac{1}{K} [\log M_{CX} + \log M_{CY} + \log M_{Y1}] 	\nonumber	
\\ & \le & I(X;Y) + h_{Y} - I(X;Y) -\epsilon_0  \label{num3.2.1}
\\ & = & h_{Y} - \epsilon_0		\nonumber
\\ & \le & R_{kX} + R_{kY} - \epsilon_0
\label{theorem2_proof_13}
\end{eqnarray}

The security levels thus result:
\begin{eqnarray}
&& \frac{1}{K} H(X^K|W_1, W_2, Y^{K_2})  \nonumber
\\ & = & \frac{1}{K} H(X^K|W_{X1} \oplus W_{kY1}, W_{X2}, W_{CX} \oplus W_{kCX}		\nonumber
\\ && W_{Y1} \oplus W_{kY1}, W_{Y2}, W_{CY} \oplus W_{kCY}, Y^{K_2})			\nonumber
\\ & \ge & \frac{1}{K} H(X^K, Y^K|W_{X1}\oplus W_{kY1}, W_{X2}, W_{Y1} \oplus W_{kY1} 
\\ && W_{Y2}, Y^{K_2}) - \epsilon_0^{''}			\nonumber
\\ & = & \frac{1}{K} H(X^K, Y^K|W_{X2}, W_{Y2}, Y^{K_2}) - \epsilon_0^{''}		\nonumber
\\ & \geq & I(X;Y) + \frac{1}{K} \log M_{X1} - \mu_C - 2\epsilon_0^{'} - \epsilon_0^{''}		\nonumber
\\ & = & I(X;Y)+ \text{min}(2^{K H(X|Y)}, 2^{h_Y - I(X;Y)}) \nonumber
\\ & - & \mu_C - 2\epsilon_0^{'} - \epsilon_0^{''}	 \nonumber
\\ & \ge & h_{Y} - \mu_C - 2\epsilon_0^{'} - \epsilon_0^{''}   \nonumber
\\ & \ge & h_{X}
\label{theorem2_proof_16}
\end{eqnarray}

\begin{eqnarray}
\frac{1}{K} H(Y^K|W_1, W_2) & = & \frac{1}{K} H(Y^K|W_{X1} \oplus W_{kX1}, \nonumber
\\ && W_{X2}, W_{CX} \oplus W_{kCX}		\nonumber
\\ && W_{Y1} \oplus W_{kY1}, W_{Y2} \nonumber
\\ && W_{CY} \oplus W_{kCY}, Y^{K_2})			\nonumber
\\ & \ge & \frac{1}{K} \log M_{Y1} + I(X;Y) - \mu_C - \mu_Y -\epsilon_0^{'} \nonumber
\\ & = & I(X;Y) + \text{min} (H(X|Y), h_{Y} - I(X;Y)) \nonumber
\\ & - & \mu_C - \mu_Y -\epsilon_0^{'}			\label{num5.44}
\\ & \ge & h_{Y} - \epsilon^{'}_0
\label{theorem2_proof_161}
\end{eqnarray}

where \eqref{num5.44} comes from \eqref{theorem1_proof_65}.

Therefore ($R_X$, $R_Y$, $R_{kX}$, $R_{kY}$, $h_{X}$, $h_{Y}$) is admissible from \eqref{theorem2_proof_11} -  \eqref{theorem2_proof_161}.

Next the case where $h_X \le I(X;Y)$ is considered. If $h_Y > I(X;Y)$ the following results.
The codewords $W_1$ and $W_2$ and their keys $W_{kX}$ and $W_{kY}$ are now defined:

\begin{eqnarray}
W_1 = (W_{X1},  W_{X2}, W_{CX} \oplus W_{kCX}) 
\label{theorem2_proof_7.1.1}
\end{eqnarray}

\begin{eqnarray}
W_2 = (W_{Y1} \oplus W_{kY1}, W_{Y2}, W_{CY} \oplus W_{kCY})
\label{theorem2_proof_8.1.1}
\end{eqnarray}

\begin{eqnarray}
W_{kX} = (W_{kCX})
\label{theorem2_proof_10.1.1}
\end{eqnarray}

\begin{eqnarray}
W_{kY} = (W_{kY1}, W_{kCY})
\label{theorem2_proof_10.1.2}
\end{eqnarray}

\begin{eqnarray}
M_{Y1} = 2^{K(h_{Y} - I(X;Y)})
\label{theorem1_proof_9.1.333}
\end{eqnarray}

where $W_\alpha \in I_{M_\alpha} = \{0, 1, \ldots, M_\alpha - 1\}$. The wiretapper will not know the $W_X$ and $W_Y$ that are covered with keys.

In this case, $R_X$, $R_Y$, $R_{kX}$ and $R_{kY}$ satisfy that

\begin{eqnarray}
\frac{1}{K} [\log M_{kX} + \log M_{kY}] & = & \frac{1}{K} [\log M_{CX} + \log M_{Y1} + \log M_{CY}]	\nonumber
\\ & \le & I(X;Y) + \frac{1}{K} \log M_{Y1} -\epsilon_0  \nonumber
\\ & = & I(X;Y) + h_{Y} - I(X;Y) - \epsilon_0  \label{num333.11}
\\ & = & h_{Y} - \epsilon_0 \nonumber
\\ & \le & R_{kX} + R_{kY} + \epsilon_0
\label{theorem2_proof_13.1.1}
\end{eqnarray}

where \eqref{num333.11} results from \eqref{theorem1_proof_9.1.333}.

The security levels thus result:
\begin{eqnarray}
&& \frac{1}{K} H(X^K|W_1, W_2) \nonumber
\\ & = &  \frac{1}{K} H(X^K|W_{X1}, W_{X2} \nonumber
\\ && W_{CX} \oplus W_{kCX}, W_{Y1} \oplus W_{kY1}, W_{Y2},		\nonumber
\\ && W_{CY} \oplus W_{kCY}, Y^{K_2})			\nonumber
\\ & \geq & I(X;Y)  - \mu_C - \epsilon_0^{'}	
\\ & = & I(X;Y) - \mu_C  - \epsilon^{'}_0	\label{num5.1.1}
\\ & \geq & h_X - \epsilon^{'}_0
\label{theorem2_proof_16.1.1}
\end{eqnarray}

where \eqref{num5.1.1} results from \eqref{theorem1_proof_9.1.333}.

\begin{eqnarray}
&&\frac{1}{K} H(Y^K|W_1, W_2) \nonumber
\\ & = & \frac{1}{K} H(Y^K|W_{X1}, W_{X2} \nonumber
\\ && W_{CX} \oplus W_{kCX}, W_{Y1} \oplus W_{kY1}, 		\nonumber
\\ && W_{Y2}, W_{CY} \oplus W_{kCY}, Y^{K_2})			\nonumber
\\ & \geq & I(X;Y) + \frac{1}{K} \log M_{Y1} \nonumber
\\ & - & \mu_C - \mu_Y - \epsilon_0	
\\ & = & I(X;Y) + h_{Y} - I(X;Y) \nonumber
\\ & - & \mu_C - \mu_Y  - \epsilon^{'}_0	\label{num5.1.11}
\\ & \ge & h_{Y} - \epsilon^{'}_0
\label{theorem2_proof_16.1.1}
\end{eqnarray}

where \eqref{num5.1.11} holds from \eqref{theorem1_proof_9.1.333}.

Next the case where $h_Y \le I(X;Y)$ is considered. 
The codewords $W_1$ and $W_2$ and their keys $W_{kX}$ and $W_{kY}$ are now defined:

\begin{eqnarray}
W_1 = (W_{X1},  W_{X2}, W_{CX} \oplus W_{kCX}) 
\label{theorem2_proof_7.1.1.1}
\end{eqnarray}

\begin{eqnarray}
W_2 = (W_{Y1}, W_{Y2}, W_{CY})
\label{theorem2_proof_8.1.1.1}
\end{eqnarray}

\begin{eqnarray}
W_{kX} = W_{kCX}
\label{theorem2_proof_10.1.1.1}
\end{eqnarray}

\begin{eqnarray}
M_{CX} = 2^{K h_Y}
\label{theorem1_proof_9.1.333.1}
\end{eqnarray}

where $W_\alpha \in I_{M_\alpha} = \{0, 1, \ldots, M_\alpha - 1\}$. The wiretapper will not know the $W_X$ and $W_Y$ that are covered with keys.

In this case, $R_X$, $R_Y$, $R_{kX}$ and $R_{kY}$ satisfy that

\begin{eqnarray}
\frac{1}{K} [\log M_{kX} + \log M_{kY}] & = & \frac{1}{K} \log M_{CX}	\nonumber
\\ & = & h_Y \label{num333.11.1}
\\ & \le & R_{kX} + R_{kY}
\label{theorem2_proof_13.1.1}
\end{eqnarray}

where \eqref{num333.11.1} results from \eqref{theorem1_proof_9.1.333.1}.

The security levels thus result:
\begin{eqnarray}
&& \frac{1}{K} H(X^K|W_1, W_2) \nonumber
\\ & = & \frac{1}{K} H(X^K|W_{X1}, W_{X2} \nonumber
\\ && W_{CX} \oplus W_{kCX}, W_{Y1}, W_{Y2},		\nonumber
\\ && W_{CY}, Y^{K_2})			\nonumber
\\ & \geq & h_Y  - \mu_C - \epsilon_0^{'}	\label{num5.1.1.1}
\\ & \geq & h_X - \epsilon^{'}_0
\label{theorem2_proof_16.1.1}
\end{eqnarray}

where \eqref{num5.1.1.1} results from \eqref{theorem1_proof_9.1.333.1}.

\begin{eqnarray}
\frac{1}{K} H(Y^K|W_1, W_2) & = & \frac{1}{K} H(Y^K|W_{X1}, W_{X2} \nonumber
\\ && W_{CX} \oplus W_{kCX}, W_{Y1}, W_{Y2},		\nonumber
\\ && W_{CY}, Y^{K_2})			\nonumber
\\ & \geq & h_Y - \mu_C - \mu_Y - \epsilon_0	\label{num5.1.11.1}	
\\ & \ge & h_{Y} - \epsilon^{'}_0
\label{theorem2_proof_16.1.1.1}
\end{eqnarray}

where \eqref{num5.1.11.1} holds from \eqref{theorem1_proof_9.1.333.1}.

Therefore ($R_X$, $R_Y$, $R_{kX}$, $R_{kY}$, $h_{X}$, $h_{Y}$) is admissible for $\text{min}(h_X, h_Y)$ \eqref{theorem2_proof_7.1.1.1} -  \eqref{theorem2_proof_16.1.1.1}.
\end{proof}

\subsection{Converse proofs for Theorems 1-2}

From Slepian-Wolf's theorem we know that the channel rate must satisfy $R_X \geq H(X|Y)$, $R_Y \geq H(Y|X)$ and $R_X + R_Y \geq H(X,Y)$ to achieve a low error probability when decoding.
Hence, only the key rates are considered in this subsection. 
\\

\textit{Converse part of Theorem 1:}
\begin{eqnarray}
R_{kX} & \geq & \frac{1}{K} \log M_{kX} - \epsilon		\nonumber
\\ & \geq & \frac{1}{K} H(W_{kX}) - \epsilon			\nonumber
\\ & \geq & \frac{1}{K} H(W_{kX}|W) - \epsilon			\nonumber
\\ & = & \frac{1}{K} [H(W_{kX}) - I(W_{kX}; W)] - \epsilon		\nonumber
\\ & = & \frac{1}{K} H(W_{kX}|X^K, Y^K, W) + I(W_{kX}; W) 		\nonumber
\\ & + & I(W_{kX};X|Y, W) + I(X, Y, W_{kX}|W) 		\nonumber
\\ & + & I(Y, W_{kX}|X, W) - I(W_{kX}; W) - \epsilon		\nonumber
\\ & = & \frac{1}{K} [H(X^K, Y^K|W) - H(X^K,Y^K|W, W_{kX})] - \epsilon		\nonumber
\\ & \geq & h_{XY} - \frac{1}{K} H(X^K,Y^K|W, W_{kX})  - \epsilon  \label{conv_1}
\\ & = & h_{XY} - \frac{1}{K} H(Y^K|X^K) -\mu_C - \epsilon  	- \epsilon_0^{''}	\nonumber
\\ & \geq & h_{XY} -\mu_C - \epsilon - \epsilon_0^{''}
\end{eqnarray}

where $W = (W_1, W_2, Y^{K_2})$ are the wiretapped portions, \eqref{conv_1} results from equation \eqref{cond8.1}. Here, we consider the extremes of $H(Y|X)$ and $H(W_Y)$ in order to determine the limit for $R_{kX}$. When this quantity is minimum then we are able to achieve the maximum bound of $h_{XY}$.

\begin{eqnarray}
R_{kY} & \geq & \frac{1}{K} \log M_{kY} - \epsilon		\nonumber
\\ & \geq & \frac{1}{K} H(W_{kY}) - \epsilon			\nonumber
\\ & \geq & \frac{1}{K} H(W_{kY|W}) - \epsilon		\nonumber
\\ & = & \frac{1}{K} [H(W_{kY}) - I(W_{kY}; W) - \epsilon		\nonumber
\\ & = & \frac{1}{K} H(W_{kY}|X, Y, W) + I(W_{kY}; W) 		\nonumber
\\ & + & I(W_{kY};X|Y, W) + I(X, Y, W_{kY}|W) 				\nonumber
\\ & + & I(Y, W_{kY}|X, W) - I(W_{kY}; W)] - \epsilon			\nonumber
\\ & = & \frac{1}{K} [H(X^K, Y^K|W) - H(X^K,Y^K|W, W_{kY})] - \epsilon		\nonumber
\\ & \geq & h_{XY} - \frac{1}{K} H(X^K,Y^K|W, W_{kY})  - \epsilon  \label{conv_2}
\\ & = & h_{XY} - \frac{1}{K}H(X^K|Y^K) -\mu_C -\mu_Y - \epsilon  - \epsilon_0^{''} 	\nonumber
\\ & \geq & h_{XY} -\mu_C -\mu_Y  - \epsilon - \epsilon_0^{''}
\end{eqnarray}

where \eqref{conv_2} results from equation \eqref{cond8.1}. Here, we consider the extremes of $H(X|Y)$ in order to determine the limit for $R_{kY}$. When this quantity is minimum then we are able to achieve the maximum bound of $h_{XY}$.

\textit{Converse part of Theorem 2:}
\begin{eqnarray}
R_{kX} & \geq & \frac{1}{K} \log M_{kX} - \epsilon		\nonumber
\\ & \geq & \frac{1}{K} H(W_{kX}) - \epsilon		\nonumber
\\ & \geq & \frac{1}{K} H(W_{kX}|W) - \epsilon		\nonumber
\\ & = & \frac{1}{K} [H(W_{kX}) - I(W_{kX}; W)] - \epsilon		\nonumber
\\ & = & \frac{1}{K} H((W_{kX}|X^K, W) + I(W_{kX}; W) 		\nonumber
\\ & + & I(X, W_{kX}|W) - I(W_{kX}; W) - \epsilon		\nonumber
\\ & \geq & \frac{1}{K} I(X^K, W_{kX}|W) - \epsilon		\nonumber
\\ & = & \frac{1}{K} [H(X^K|W) - H(X^K|W, W_{kX})] - \epsilon	\nonumber
\\ & \geq & h_{X} - H(W_{CY}) -\mu_C - \epsilon - \epsilon_0^{''}  \label{conv_3}
\\ & \geq & h_{X}  - \mu_C -  \epsilon - \epsilon_0^{''}
\end{eqnarray}

where $W= (W_1, W_2)$, \eqref{conv_3} results from \eqref{cond7}. The consideration here is that $H(Y^{K_2})$ represents the preexisting information known to an eavesdropper as an extreme case scenario.
Here, we can also consider the extremes of $H(W_{CY})$ in order to determine the limit for $R_{kX}$. When this quantity is minimum then we are able to achieve the maximum bound of $h_{X}$.

\begin{eqnarray}
R_{kY} & \geq & \frac{1}{K} \log M_{kY} - \epsilon	\nonumber
\\ & \geq & \frac{1}{K} H(W_{kY}) - \epsilon		\nonumber
\\ & \geq & \frac{1}{K} H(W_{kY}|W) - \epsilon		\nonumber
\\ & = & \frac{1}{K} [H(W_{kY}) - I(W_{kY}; W)] - \epsilon		\nonumber
\\ & = & \frac{1}{K} H(W_{kY}|Y^K, W) + I(W_{kY}; W) 		\nonumber
\\ & + & I(X, W_{kY}|W) - I(W_{kY}; W) - \epsilon		\nonumber
\\ & \geq & \frac{1}{K} I(Y^K, W_{kY}|W) - \epsilon				\nonumber
\\ & = & \frac{1}{K} [H(Y^K|W) - H(Y^K|W, W_{kY})] - \epsilon		\nonumber
\\ & \geq & h_{Y} - H(W_{CX}) -\mu_C -\mu_Y - \epsilon - \epsilon_0^{''} \label{conv_4}
\\ & \geq & h_{Y} -\mu_C -\mu_Y - \epsilon - \epsilon_0^{''}
\end{eqnarray}

where \eqref{conv_3} results from \eqref{cond8}. The same consideration as above for $ H(Y^{K_2})$ is presented here. Here, we consider the extremes of $H(W_{CX})$ in order to determine the limit for $R_{kY}$. When this quantity is minimum then we are able to achieve the maximum bound of $h_{Y}$.

\section{Conclusion}
A Shannon cipher approach for two correlated sources with compromised source symbols across a channel with an eavesdropper has been detailed. This is a more generalised approach of Yamamoto's~\cite{shannon1_yamamoto} model and has made use of the idea of wiretapping source bits from Luo \textit{et al.} \cite{ref14_luo_mitpant}. From Section II it has been noticed that the model can be implemented for more than two sources with Shannon's cipher system. The channel and key rates to achieve perfect secrecy have been provided, where the rates have been quantified in terms of bounds for the channel and key rates. A method to reduce the key length via masking using bits that were not eavesdropped was also detailed. 

\bibliographystyle{IEEEtran}	% (uses file "IEEEtran.bst")
\bibliography{bib}

\end{document}